\documentstyle[aps,twocolumn,prl,epsf]{revtex}

\begin{document}
\draft

\twocolumn[\hsize\textwidth\columnwidth\hsize\csname @twocolumnfalse\endcsname
\title{Dynamics of disordered heavy Fermion systems}
\author{A.\ Chattopadhyay and M.\ Jarrell}
\address{ Department of Physics, University of Cincinnati 
\\Cincinnati, Ohio 45221
}
\date{\today}
\maketitle

\widetext
\begin{abstract}
\noindent
Dynamics of the disordered heavy Fermion model of Dobrosavljevi\'{c} 
et al.\ are calculated using an expression for the spectral function 
of the Anderson model which is consistent with quantum Monte Carlo 
results.  We compute $\Sigma(\omega)$ for three distributions of Kondo 
scales including the distribution of Bernal et al.\ for UCu$_{5-x}$Pd$_{x}$. 
The corresponding low temperature optical conductivity shows a 
low-frequency pseudogap, a negative optical mass enhancement, and
a linear in frequency transport scattering rate, consistent with
results in Y$_{1-x}$U$_{x}$Pd$_{3}$ and UCu$_{5-x}$Pd$_{x}$.
\end{abstract}

]

\narrowtext
\paragraph*{Introduction.}
Over the past four decades, the Fermi-Liquid paradigm has been the 
key to our understanding of metallic behavior.  In Fermi-liquid 
theory we assume a 1:1 correspondence between the low-lying 
eigenstates of the interacting system to those of the noninteracting 
electron gas. This leads to a magnetic susceptibility that is weakly 
temperature dependent, a specific heat linear in T, and a low 
temperature resistivity that is quadratic. 
\par\indent However, recently there has been a 
great deal of experimental interest in disordered heavy-Fermion compounds\cite{hfnfl} (e.g.Y$_{1-x}$U$_{x}$Pd$_{3}$, UCu$_{5-x}$Pd$_{x}$), 
where strong electronic correlations preclude Fermi-liquid 
behavior.  The non Fermi-liquid behavior in these compounds 
is characterized by a linear resistivity at low T, a logarithmic low 
temperature divergence of the susceptibility and the specific heat 
coefficient, an optical conductivity with a low frequency 
pseudogap and a linear transport scattering rate at low frequencies
\cite{hfnfl,degiorgi,vonL}.  Several models have been proposed to 
explain these experimental results. Among them are theories based upon 
proximity to a zero temperature quantum critical point\cite{tsvelik} and
those which explain the impurity to lattice crossover effects in 
the multi-channel Kondo model\cite{cox}.
\par\indent A few years ago, 
Dobrosavljevi\'{c} et al.\cite{disknfl} investigated a system of 
dilute magnetic impurities in 
a disordered metal. Since disorder can give rise to a 
distribution of the local density of states of the conduction 
electrons, a distribution of Kondo scales $P(T_{K})$ is induced 
that could be singular enough to produce $\chi$ and $\gamma$ that diverge 
as $T\rightarrow 0$, a strongly non-Fermi-liquid behavior. More recently, 
Bernal et al.\cite{bernal} have shown that this $P(T_{K})$ is not  
appropriate for the $U$-doped heavy-Fermion systems. They propose an 
alternative spread of Kondo scales, and use it to calculate the
thermodynamics of these alloys.  
Miranda, et al.\cite{miranda} show that this distribution is consistent
with a linear in $T$ low temperature resistivity.   

\paragraph*{Formalism.} In this paper, we stay within this 
phenomenological framework and calculate dynamical 
quantities such as the optical conductivity and self-energy. 
We begin by concentrating on the Kondo regime of 
the Anderson impurity model, where d-electrons occupy the conduction 
band, and f-electrons provide the magnetic impurities for the spin-spin scattering processes.
It is known that in the low temperature limit this model corresponds
precisely to the Fermi-liquid picture of Landau, containing quasi-particle excitations from a ground state with relatively weak inter 
quasi-particle interactions ($\sim k_{B}T_{K}$). The electronic density of states has a resonance at the Fermi level, giving significant impurity contributions to the specific heat and magnetic susceptibility. Following Doniach and Sunjic\cite{ds}, Frota and Oliveira\cite{oliveira} argued that
the Doniach-Sunjic form, modified to account for the $\pi/2$ phase shift, 
should describe the shape of Kondo resonance.  Their expression is in
agreement with their results for the Kondo resonance
obtained from numerical renormalization group calculations\cite{hrk}, 
as well as low-temperature quantum Monte Carlo (QMC) results analytically 
continued with the Maximum Entropy method (MEM)\cite{silver,jarrell_mem}. 
\par\indent We can generalize their expression to finite temperatures and 
get
\begin{equation}
A_f(\omega,T_{K})= \frac{1}{\pi\Delta}
{\rm{Re}}
\left[
i\Gamma_K/(\omega+i\sqrt{\Gamma_K^2+\gamma^2})
\right]^{1/2}\,,
\end{equation}
where $\Delta$ is the f-d hybridization energy, and $\Gamma_{K}=(\pi/2)^{2}T_{K}$ 
is the half-width of this resonance, $T_{K}$ being the Kondo 
temperature.  We add a temperature dependent width $\gamma$ to the
original expression of Frota and Oliveira\cite{oliveira}, with $\gamma$ determined by fitting to QMC-MEM results\cite{silver}.  As shown in 
Fig.[1], this form continues to exhibit remarkable agreement with the 
shape of the Kondo
resonance (the low-frequency peak) obtained from
QMC-MEM, {\em{even at finite temperatures}}.  By comparison 
with a wide range of Anderson impurity spectra, we were able to
obtain the universal function $\gamma(T/T_K)$ for over three
decades of $T/T_K$. For $T/T_K\leq0.3$, we use $\gamma(T)=4.52 T$,
a value derived from Nozi\'{e}res'\cite{nozier} phenomenological 
Fermi-liquid description of the Kondo problem at low temperatures. 
The result for $\gamma(T/T_{K})$ will be presented in a table of an
upcoming publication.

	The Hilbert transform of $A_f(\omega,T_{K})$ then gives
the average impurity t-matrix
\begin{equation}
t_f(z)= \int dT_K P(T_K) \int d\omega \frac{V^2 A_f(\omega,T_{K})}{z-\omega}\,.
\end{equation}
where $V$ is the f-d hybridization.  Following Miranda, et al.\cite{miranda} 
we use a dynamical mean-field approximation\cite{metzvoll}, which becomes 
exact in the limit of infinite dimensions, to calculate the lattice self 
energy from a concentration $x$ of substitutional Kondo impurities
\begin{equation}
\Sigma(\omega)=\frac{x t_f(\omega)}{1+ x t_f(\omega){\cal{G}}(\omega)}\,,
\end{equation}
where ${\cal{G}}(\omega)$ describes the average effective medium
of the impurity.  It is related to the average local host greens function
$G$ 
\begin{equation}
G(\omega) = \int d\epsilon \frac{N(\epsilon)}{\omega-\epsilon +\mu -\Sigma(\omega)}\,,
\end{equation}
through the relation 
\begin{equation}
{\cal{G}}^{-1} = G^{-1} + \Sigma\,,
\end{equation}
where $N(\epsilon) = \frac{1}{t^{*}\sqrt\pi} e^{-\epsilon^{2}/t^{*^{2}}}$
and we set $t^{*} = 10,000K$ to establish a unit of energy and temperature. 
The solutions of Eqns.\ 1--5 then give the full lattice self energy.
\par\indent The knowledge of this self energy enables one to calculate
physical quantities\cite{pruschke} like transport coefficients and the 
optical conductivity. In this paper we concentrate on the optical 
conductivity $\sigma(\omega)$.  
It is measured in units of $\sigma_{0}= e^2\pi/2\hbar a$, 
which with $h/e^2 \approx 2.6\cdot10^4\Omega$, varies between 
$10^{-3}...10^{-2}[(\mu\Omega cm)^{-1}]$, depending on the lattice 
constant $a$. 
\paragraph*{Results}  We choose three different distributions $P(T_K)$,
corresponding to strong\cite{disknfl}, weak\cite{bernal} and a
phenomenological disorder motivated by Miranda et al.'s\cite{miranda} 
argument that the experimental $P(T_K)$ should be
relatively constant at low temperatures. Also consistent with Miranda
we consider the distribution of Kondo scales as arising from a 
distribution of
couplings between the conduction and the f-electron spins $P(J)$. 
For the strongly disordered sample, $P(T_K)$ has the form\cite{disknfl}
\begin{eqnarray}
P_{sd}(T_{K}) & = & (4\pi)^{-1/2}\frac{1}{T_{K}{\rm{ln}}(t^*/T_{K})}\times \nonumber\\
  & &  {\rm{exp}}\{-0.25{\rm{ln}}^{2}[0.217e^{-1}{\rm{ln}}
(t^*/T_{K})]\}\,,
\end{eqnarray}
where we have used a bulk Kondo temperature $T_{K}^{0}= 10^{2}K$ 
(in most U-based HF systems  $T_{K}^{0}= 10^{2}K$  varies between 
100-200 K).  The weak disorder is characterized by a Gaussian distribution $P(J)$ of width $2u = 0.01$, where $u$ is a disorder parameter
\cite{disknfl} and a higher value of $u$ corresponds to more disorder. 
This leads to  
\begin{eqnarray}
P_{wd}(T_{K}) & = & 
\frac{1}{\sqrt{0.01\pi}}  \frac{1}{0.217T_{K}{\rm{ln}}^{2}(T_{K}/t^*)}\times \nonumber\\
  & &  \exp\left\{-100\left[\frac{1}{0.217{\rm{ln}}(T_{K}/t^*)}+1\right]^{2}\right\}\,.
\end{eqnarray}
For the phenomenological spread of Kondo scales, we assume the form 
\begin{equation}
P_{ph}(T_{K}) = \frac{0.01}{e^{(T_{K}-T_{K}^{0})}+1}\,.
\end{equation}
This distribution is not based on microscopics; it simply satisfies the experimental criterion of constancy at low $T_{K}$ and looks qualitatively similar to the $P(T_K)$ for UCu$_{5-x}$Pd$_x$\cite{miranda}. Unlike the $1/(T_K{\rm{ln}}T_{K})$ divergence of the strongly disordered case, $P_{ph}(T_K)$ has a finite number of spins with $T_{K} = 0$. 
\par\indent The nature of the disorder gives 
rise to different physics in each case, as is manifested in the 
functional form of ${\rm{Im}}\Sigma(\omega)$ [Fig.2] (these plots are 
for $T=0$). For a weakly disordered system, ${\rm{Im}}\Sigma(\omega)$
has the form $-c + \omega^2$ as $\omega\rightarrow0$ where $c$ is a 
constant\cite{caveat2}. This suggests a finite lifetime for the 
electrons at the Fermi energy at zero temperature which makes it 
different from a normal (pure metal) Fermi liquid. 
But since there are no unquenched spins at $T=0$, the system 
does form a local Fermi liquid, with a resistivity $\rho(T) \sim \rho(0) - AT^2$. 
For the case of strong disorder, $P_{sd}(T_K)$ is divergent at
low $T_K$.  Even though we are on the metallic side of the metal-insulator transition, a large number of spins with very low Kondo temperatures 
gives a non Fermi-liquid ground state. 
For this scenario, we find that ${\rm{Im}}\Sigma(\omega) 
\sim -c+\omega^{1/4}$ at low $\omega$.

$P_{ph}(T_K)$, the distribution of Kondo scales that is 
relevant to UCu$_{5-x}$Pd$_{x}$, is intermediate 
between these two cases. It gives us an ${\rm{Im}}\Sigma(\omega)$ that
behaves like $-c+\left|\omega\right|$ as $\omega\rightarrow0$. We 
think that this behavior of ${\rm{Im}}\Sigma(\omega)$ can be
understood through the following simple argument. Oliveira's 
expression for the f-electron spectral function at $T=0$ can be 
expanded near $\omega=0$ to give a form 
$A_f(\omega,T_{K}) \sim 1 - \alpha(\omega/T_{K})^2$, where 
$\alpha$ is a constant. Given that $P_{ph}(T_K) \approx constant$ at 
low $T_K$, $\int_{0}^{\infty}A_f(\omega,T_{K})\,P_{ph}(T_K)\,dT_{K}$
gets its dominant contribution from the region where 
$T_{K} \geq \omega$. 
If we change the lower limit of the integral from $0$ to $\omega$ and
make use of the fact that $P_{ph}(T_K)$ has a finite upper cutoff 
($\sim 100K$),
${\rm{Im}}\Sigma(\omega)$ turns out to be 
 $-c+\left|\omega\right|$. In calculating the conductivity, the energy
$\omega$ of the electron is averaged over a region of width $k_{B}T$ 
near the Fermi surface. This replaces $\left|\omega\right|$ by $T$, 
giving a resistivity linear in temperature\cite{miranda} which is 
observed experimentally in Y$_{1-x}$U$_x$Pd$_3$ and UCu$_{3.5}$Pd$_{1.5}$\cite{hfnfl}.This result is confirmed by 
direct calculation of the resistivity (not shown).

   Fig.[3] shows the temperature dependence of the optical conductivity of 
the phenomenologically disordered system. Consistent with what is 
seen in Y$_{1-x}$U$_x$Pd$_3$ and UCu$_{3.5}$Pd$_{1.5}$\cite{degiorgi}, 
the Drude-like peak is recovered when $T \agt 100K$, since spins 
with essentially all possible values of $T_{K}$ are participating 
in the dynamics at this temperature.  The inset lists the zero 
temperature optical conductivities for the three distribution of 
Kondo scales.  All are characterized by a vanishing Drude weight at 
low $T$, along with a finite frequency peak. As the disorder is increased, 
this peak moves towards lower frequencies, concomitant with the decrease 
in the average value of $T_{K}$ (which is different than $T_{K}^{0}$). 
The Drude peak at $T=0$ is recovered from the weak disorder case if 
$u\rightarrow{0}$ in $P(J)$, giving 
$P_{wd}(T_{K}) \propto \delta(T_{K}-T_{K}^{0})$ ($T_{K}^{0}$ is
the bulk value), which takes us to the single Kondo scale physics. 

The optical conductivity of metals, even non Fermi liquid metals, is
usually analyzed by rewriting it in a generalized Drude form
\cite{degiorgi}
\begin{equation}
\sigma(\omega) = \frac{\omega_{p}^{2}}{4\pi} \frac{1}{{\Gamma(\omega)} - i{\omega}(1+\lambda(\omega))}\,,
\end{equation}
where $\sigma(\omega)=\sigma_1(\omega)+i\sigma_2(\omega)$\cite{caveat}.
We calculate the transport relaxation rate $\Gamma(\omega)$ and 
the optical mass enhancement $\left(1+\lambda(\omega)\right)$ for the 
three distributions of Kondo scales.  However, as shown in Fig.~4(a),
only the phenomenological distribution results in a linear in frequency
zero temperature $\Gamma(\omega)$, consistent with what is seen in
Y$_{1-x}$U$_x$Pd$_3$ and UCu$_{3.5}$Pd$_{1.5}$\cite{degiorgi}.
In each case, we find that the low frequency optical mass enhancement 
$1+\lambda(0)$ is also negative (however, in the case of 
weak disorder a positive mass is recovered as $u\to 0$).  

$\Gamma(\omega)$ for the phenomenological distribution at several 
different temperatures is plotted in Fig.~4(b). Here, consistent 
with Degiorgi, et al.\cite{degiorgi} we fit the relaxation 
rate to $\Gamma(\omega)=\Gamma_0\left(1-\left(T/T_0\right)^n -
\left(\omega/\omega_0\right)^n\right)$ in the low frequency region.
We find that $T_0\approx 85K$ and $\omega_0\approx0.09t^{*}$, and are
roughly constant in temperature.  At high temperatures, $\Gamma(\omega)$ 
and $1+\lambda(\omega)$ (Fig.~4(c)) are weakly frequency 
dependent, $1+\lambda(\omega)>0$ and $n=2.00$, as expected 
for  a Fermi liquid.  Thus, as seen in Fig.~3, a Drude peak is recovered in 
$\sigma_1(\omega)$.  As the temperature is lowered, $\Gamma(\omega)$ 
and $1+\lambda(\omega)$ become strongly frequency dependent, 
$1+\lambda(0)<0$, and $n \approx 1$, features which we
believe should be viewed as characteristic of a non-Fermi liquid.  
A very similar sequence of features are seen in Y$_{1-x}$U$_x$Pd$_3$ 
and UCu$_{3.5}$Pd$_{1.5}$\cite{degiorgi}.

\par\indent Within the formalism developed thus far, we can also calculate
the magnetic susceptibility $\chi(T)$ by disorder averaging Krishnamurthy's
\cite{hrk} universal susceptibility for a single Kondo scale. It is 
fairly straight-forward to compute the spin-relaxation rate in
NMR ($1/T_{1}$) as well. These quantities will be addressed in
an upcoming publication.
   
\paragraph*{Conclusion.} Within the Kondo disorder 
 model\cite{disknfl,bernal,miranda},
we calculate dynamics for some U-based heavy Fermion systems.  We 
observe a linear resistivity at low $T$ consistent with Ref.~[8], 
the lack of a Drude peak and a low-frequency pseudogap in the real 
part of the optical conductivity, a negative low temperature optical 
mass, and a linear in frequency
optical dynamical scattering rate.  All these features are observed
in Y$_{1-x}$U$_x$Pd$_3$ and UCu$_{3.5}$Pd$_{1.5}$\cite{hfnfl,degiorgi}.
Thus, we conclude that the phenomenological distribution of Kondo
scales model is sufficient to describe the dynamics of these
disordered systems.  It is important to stress that Kondo disorder is not 
the sole possible explanation of non Fermi-liquid behavior in these systems
\cite{tsvelik,cox}.  In fact, it has recently been shown that the two-channel
Kondo lattice model displays remarkably similar optical properties\cite{sces}.
However, it remains to be seen if an appropriate two-channel Kondo
model can accurately describe the transport and optical properties of 
{\em{dilute}} systems such as Y$_{1-x}$U$_x$Pd$_3$.

\par\indent We are most grateful to Leo Degiorgi for his 
careful and critical reading of the manuscript. His comments led
to several corrections and clarifications.
We would also like to acknowledge useful discussions with 
D.L.\ Cox,
B.\ Goodman,
M.\ Ma,
Anirvan\ Sengupta,
R.\ Serota, and
Axel\ Zibold.
This work was supported by NSF grants DMR-9406678 and DMR-9357199.  
Computer support was provided by the Ohio Supercomputer Center.

\begin{figure}
\epsfxsize=3.0in
\epsffile{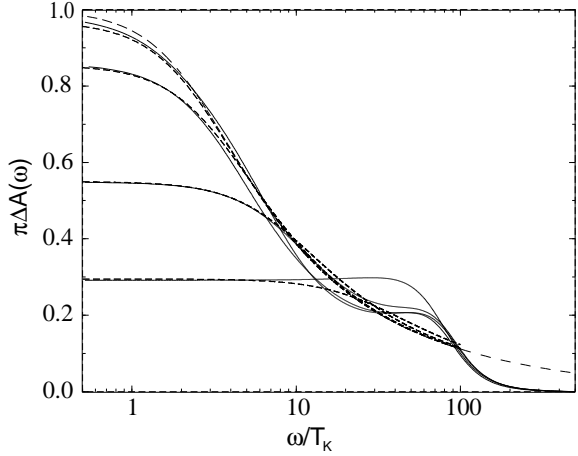}
\caption[]{{Spectral function for the f-electron in the Kondo limit of
the Anderson model.  The long-dashed line is the Doniach-Sunjic result\cite{oliveira} (true for $T=0$). The dotted lines represent 
the extension of the Doniach-Sunjic result to finite $T$ which is used for calculating the dynamics in this paper; the solid lines represent QMC-MEM results. $T/T_{K}$ = 0.2, 0.8, 3.2 and 12.8 for the curves from top 
to bottom. For $T/T_{K}\leq{0.3}$, $A_f(\omega,T_{k})$ has a width $\gamma=4.52T$ which comes from Fermi-liquid theory\cite{nozier}. 
For higher $T/T_{K}$, we adjust $\gamma$ to fit the QMC-MEM data. }}     
\end{figure}

\begin{figure}[t]
\epsfxsize=3.8in
\epsfysize=2.5in
\epsffile{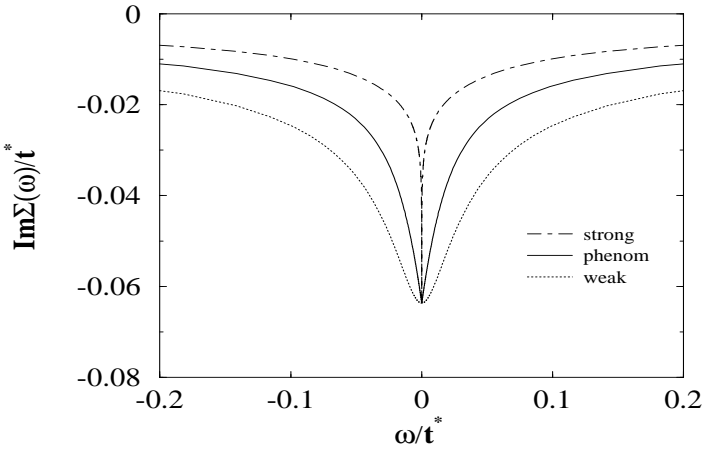}
\caption[]{Imaginary part of the conduction electron self energy when $T=0$
and $x=0.2$. The three different curves denote different distributions of disorder. The lowest one corresponds to very weak disorder 
($P(T_{K})\rightarrow {0}$ as $T_{K}\rightarrow {0}$) and at 
low $\omega$ has the form ${\rm{Im}}\Sigma(\omega)\propto -c+\omega^2$, 
giving a local Fermi liquid with $\rho(T) = \rho(0) - AT^{2}$. 
The highest one is for strong disorder (still metallic), with a 
$\omega^{1/4}$ dependence as $\omega\rightarrow0$. The one in the 
middle corresponds to a phenomenological distribution of Kondo scales 
suitable for the heavy-fermion systems.  It approaches the form 
${\rm{Im}}\Sigma(\omega)\propto -c+\left|\omega\right|$ as $\omega\rightarrow0$. This strongly hints towards a linear resistivity 
at low $T$ in these compounds. } 
\end{figure}   

\begin{figure}[t]
\epsfxsize=3.8in
\epsfysize=2.5in
\epsffile{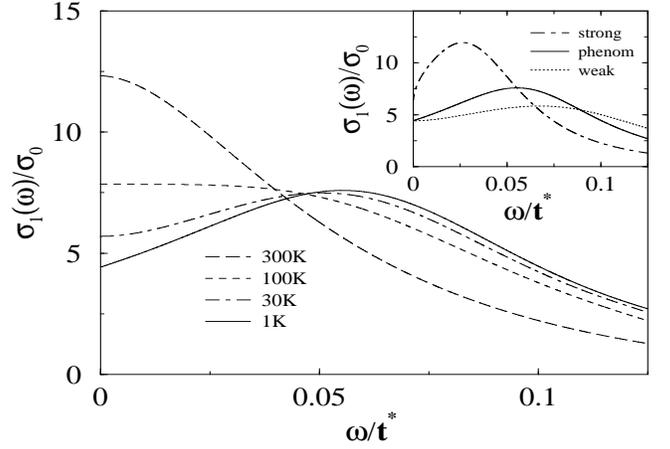}
\caption{Optical conductivity for the phenomenological distribution 
of Kondo scales at a few temperatures when $x=0.2$. At very low 
temperatures a finite number of unquenched spins preempt the formation
of a Fermi-liquid. The interesting feature is the development of a 
Drude peak as we go from temperatures much below the bulk Kondo value
($T_K^{0}\approx 100K$) to those much above it. The inset shows the
conductivity for the three different $P(T_{K})$ when $T=0$.  The absence 
of a Drude peak is conspicuous in all three cases.  }
\end{figure}

\begin{figure}[t]
\epsfxsize=3.8in
\epsfysize=2.5in
\epsffile{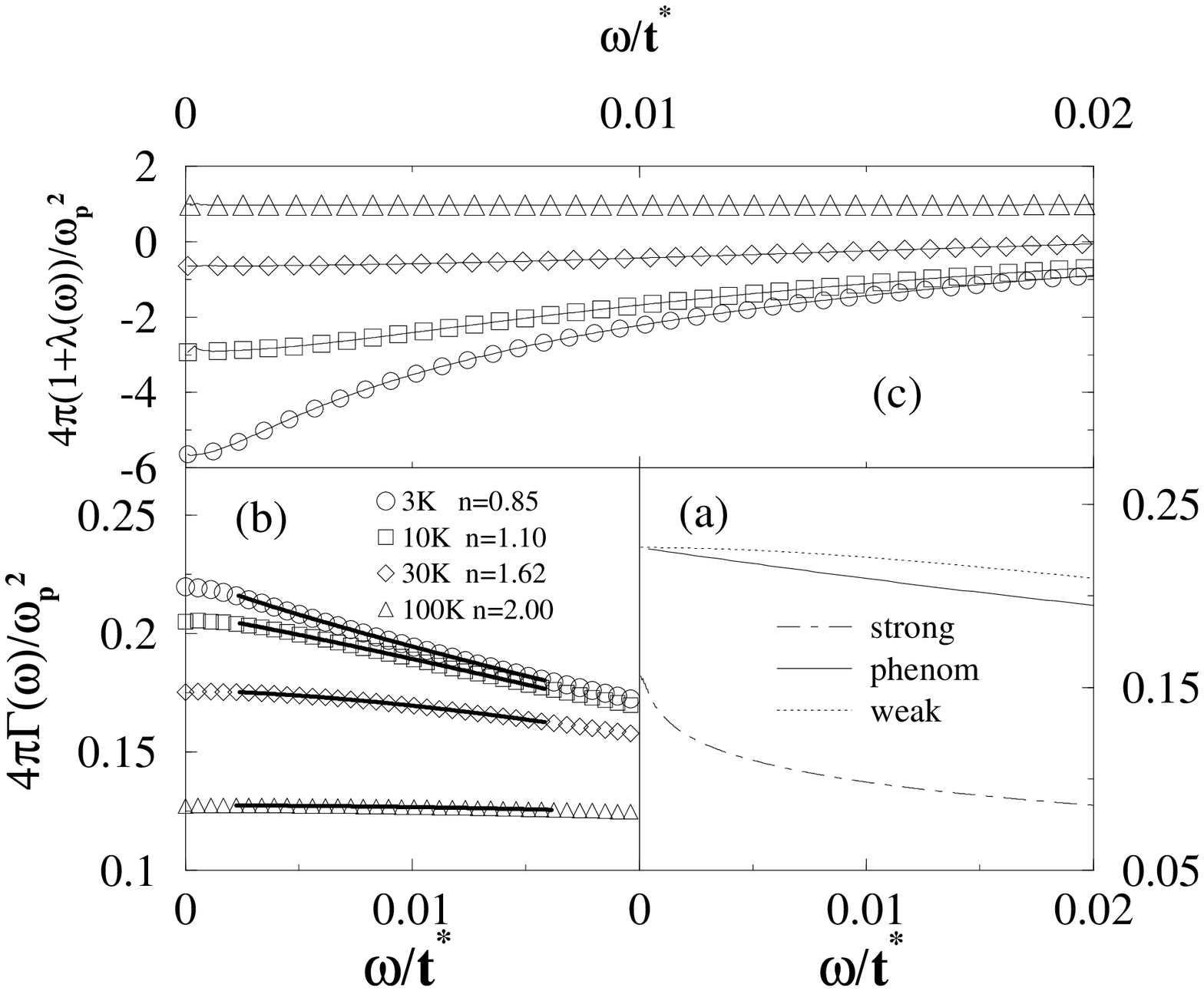}
\caption[]{{(a) Frequency dependence of the scattering relaxation rate
$\Gamma(\omega)$ at $T={0}$ for the three different $P(T_{K})$ when
$x=0.2$.  Note that $\Gamma(\omega)$ is linear in $\omega$, and therefore consistent with experiment\cite{degiorgi}, only for the phenomenological distribution of Kondo scales. $\Gamma(\omega)$ corresponding to
weak disorder has an $\omega^2$ behavior as $\omega\to 0$, suggesting
the formation of a local Fermi liquid. (b) $\Gamma(\omega)$ for 
$P_{ph}(T_{K})$ at different temperatures; in each case the solid line 
is a fit to the form $\Gamma(\omega)=\Gamma_0\left(1-\left(T/T_0\right)^n -
\left(\omega/\omega_0\right)^n\right)$. We see that up until $10K$, 
we have a scattering rate that is roughly linearly decreasing in $\omega$
and $T$. (c) Optical mass enhancement for $P_{ph}(T_{K})$ at different 
temperatures (symbols are the same as (b)).  At low temperatures 
$1+\lambda(0)<0$ indicative of a non-Fermi liquid. }}
\end{figure}


\begin{references}
\bibitem{caveat} Given the nonanalytic nature of the self energy, and thus, 
presumably the dynamic current-current susceptibility, it is important to
calculate $\sigma_2(\omega)$ exactly the same way that the experimentalists
do, using a Hilbert transform of $\sigma_1(\omega)$ (rather than the
dynamic current-current susceptibility, $-i\omega \sigma(\omega)$.)
\bibitem{caveat2} Strictly speaking, since $P_{wd}(T_{K})$
has a weak divergence like $1/(T_{K}ln^{2}T_{K})$ as $T_{K}\rightarrow0$,
the corresponding ${\rm{Im}}\Sigma(\omega)$ also will contain a term like 
$\omega^{\alpha}$ with $\alpha<1$.  However, the prefactor of this term is 
proportional to the relative weight of this singularity, which for our choice
of parameters is about $10^{-7}$.  Hence, the term with the fractional power 
gets suppressed by the quadratic term for all experimentally accessible
temperatures and frequencies.
\end{references}
\end{document}